\documentclass[prl,aps,amsmath,amssymb,eshowkeys,letter,twocolumn]{revtex4-2}

\usepackage{verbatim,graphics,graphicx,color,slashed,textcomp,bbm,mathdots,multirow,array}
\usepackage[normalem]{ulem} 
\usepackage[colorlinks=true,linkcolor=red,urlcolor=blue,citecolor=blue]{hyperref}
\usepackage[T1]{fontenc}
\usepackage{array}
\usepackage{booktabs}
\usepackage{autobreak}
\usepackage{float}
\usepackage{graphicx}
\usepackage{caption,subcaption}
\graphicspath{{figures/}}

\makeatletter

\newcommand{\Rmnum}[1]{\expandafter\@slowromancap\romannumeral #1@}
\makeatother

\usepackage[colorlinks=true,linkcolor=red,urlcolor=blue,citecolor=blue]{hyperref}

\graphicspath{{figures/}}

\begin{document}


\title{Detecting Ultralight Dark Matter Gravitationally \\ with Laser Interferometers in Space}
\author{Jiang-Chuan Yu$^{a,c}$}

\author{Yan Cao$^{a}$}

\author{Yong Tang$^{a,b,c,d}$}
\author{Yue-Liang Wu$^{a,b,c,e}$}
\affiliation{\begin{footnotesize}
		${}^a$University of Chinese Academy of Sciences (UCAS), Beijing 100049, China\\
		${}^b$School of Fundamental Physics and Mathematical Sciences, \\
		Hangzhou Institute for Advanced Study, UCAS, Hangzhou 310024, China \\
		${}^c$International Center for Theoretical Physics Asia-Pacific, Beijing/Hangzhou, China \\
		${}^d$National Astronomical Observatories, Chinese Academy of Sciences, Beijing 100101, China\\
		${}^e$Institute of Theoretical Physics, Chinese Academy of Sciences, Beijing 100190, China
		\end{footnotesize}}

\date{\today}

\begin{abstract}
Ultralight dark matter~(ULDM) is one of the leading well-motivated dark matter candidates, predicted in many theories beyond the standard model of particle physics and cosmology. There have been increasing interests in searching for ULDM in physical and astronomical experiments, mostly assuming there are additional interactions other than gravity between ULDM and normal matter. Here we demonstrate that even if ULDM has only gravitational interaction, it shall induce gravitational perturbations in solar system that may be large enough to cause detectable signals in future gravitational-wave (GW) laser interferometers in space. We investigate the sensitivities of Michelson time-delay interferometer to ULDM of various spins, and show vector ULDM with mass $m\lesssim 10^{-18}~$eV can be probed by space-based GW detectors aiming at $\mu$Hz frequencies. Our findings exhibit that GW detectors may directly probe ULDM in some mass ranges that otherwise are challenging to examine. 

\end{abstract}

\maketitle
\newpage

\noindent {\bf Introduction.}---%
Dark Matter (DM) constitutes the majority of matter in the universe, with supporting evidence from galactic rotational curves, bullet cluster, large-scale structure and cosmic microwave background. However, DM's identity remains unclear because all the confirmed evidence merely suggests DM must have gravitational interaction. Among the leading candidates, ultralight dark matter (ULDM) with mass $\lesssim 10^{-6}~$eV stands out attractively since it may connect to fundamental physics at very high energy~\cite{PQ1977, Weinberg1978, Wilczek1978, PhysRevD.95.043541, Marsh:2015xka, Agrawal:2021dbo}, including quantum gravity~\cite{Polchinski:1998rr, Wu:2022aet}, grand unified theory~\cite{Kim:1986ax, Fayet:1990wx, Arias:2012az}, inflation~\cite{Lyth:1998xn,Graham:2015rva, Ema:2019yrd, Ahmed:2020fhc,Kolb:2020fwh}, etc. 

There have been vast efforts and significant progresses in searching for ULDM, either presuming there are additional nongravitational interactions between ULDM and normal matter~\cite{Graham:2013gfa, Graham:2015ouw, Safronova:2017xyt} or just gravity. For instance, searching for axion-like ULDM~\cite{Preskill:1982cy, Abbott:1982af, Dine:1982ah} involves interactions with photon~\cite{VanTilburg:2015oza, Chen:2019fsq, Li:2023qyr, Liang:2018mqm, PhysRevD.103.076018, Liu:2021zlt}, electron~\cite{Stadnik:2014tta, PandaX:2017ock, XENON:2022ltv, Chao:2023dwc}, nucleon~\cite{Budker:2013hfa, Hamaguchi:2018oqw, Garcon:2019inh, Centers:2019dyn, Wei:2023rzs} or neutrino~\cite{Ge:2019tdi, Choi:2019zxy, Chen:2023vkq}. Dark photon ULDM searches assume its mixing with photon~\cite{Caputo:2021eaa, Chaudhuri:2014dla, An:2022hhb, An:2023wij, An:2024kls, Chen:2022quj, Chen:2023swh, Chen:2024aqf, Tang:2023oid, Chao:2023lox}. After the detection of gravitational waves (GWs), it was shown that gravitational-wave (GW) experiments can also be very sensitive to some of such interactions~\cite{PhysRevLett.121.061102, PhysRevD.100.123512, Guo:2019qgs, fukusumi2023upper, PhysRevD.108.083007}.


Astrophysical methods that only employ ULDM's gravitational interaction are also pushed forward, including probing metric perturbation by pulsar timing array (PTA)~\cite{Khmelnitsky:2013lxt, Porayko:2014rfa, Aoki:2016kwl, Nomura:2019cvc, Sun:2021yra, Wu:2023dnp, Guo:2023gfc, Cai:2024thd, PPTA:2021uzb, Kim:2023pkx, Kim:2023kyy}, binary dynamics~\cite{Pitjev_2013, Blas:2016ddr, Blas:2019hxz, Yuan:2022nmu}, Lyman-$\alpha$ constraints~\cite{Armengaud:2017nkf, lymanalpha2017, Kobayashi:2017jcf}, 21cm absorption line~\cite{Schneider:2018xba, Lidz:2018fqo}, superradiance around black holes~\cite{Arvanitaki:2010sy, Brito:2015oca, Arvanitaki:2016qwi, Brito:2017zvb, Davoudiasl:2019nlo, Cao:2023fyv}, and galactic dynamics~\cite{Bar:2018acw, Marsh:2018zyw, Bar:2019bqz, Safarzadeh:2019sre}. Such searches can provide complementary information and are largely model-independent unless non-gravitational interaction dramatically change the physical contexts. 

In this work, we present the searches for local ULDM of various spins by space-based GW laser interferometers. Unlike the scalar case that suffers from velocity suppression, we demonstrate that the gravitational tensor perturbations in solar system caused by vector ULDM can lead to detectable signals, which may be probed by future GW laser inteferomerters in space with arm lengths comparable to the diameter of Mars' orbit. Detection of such a signal would directly suggest the wave nature of DM and its mass range, signifying the fundamental theory of DM. 

\noindent {\bf Metric perturbations and signal response.}---%
Since we are considering the effects of ULDM in solar system at a time scale of years, we can neglect the cosmic expansion and write the metric with linear perturbations over a flat background, defining scalar metric perturbations $\Psi$ and $\Phi$, and tensor $h_{ij}$ by
\begin{equation}
d s^2=-(1+2 \Psi) d t^2+[(1-2 \Phi) \delta_{i j}+h_{i j}]d x^i d x^j.
\end{equation}
In the Poisson gauge we have $\partial_i h_{ij}=\delta_{ij}h_{ij}=0$. The energy-momentum tensor $T_{\mu \nu}$ of ULDM would source metric perturbations through the Einstein equations. For ULDM with mass $m$, energy density $\rho$ and velocity $v$, we can estimate the amplitudes of perturbations,
\begin{align}
\Psi^j \simeq \Phi^j \simeq \pi G \frac{\rho}{m^2} =& \frac{7\times10^{-26}\rho}{0.4\,\text{GeV/cm}^{3}}\left(\frac{10^{-18}\text{eV}}{m}\right)^{2},\\
h^{v}_{ij} \propto h_0\simeq \frac{8}{3}\pi G \frac{\rho}{m^2} = &\frac{2\times10^{-25}\rho}{0.4\,\text{GeV/cm}^{3}}\left(\frac{10^{-18}\text{eV}}{m}\right)^{2}, 
\end{align}
and $h^{s}_{ij}\simeq h_0 v^2/2$. Here the superscript $j=s$ and $v$ stands for contribution from scalar and vector ULDM~\footnote{We find tensor ULDM has similar behaviors as scalar ULDM.}. Note that $h^{s}_{ij}$ due to scalar ULDM are suppressed by $v^2$ (see Supplemental Material). However, the tensor perturbation $h^{v}_{ij}$ for vector ULDM is not suppressed, which is crucial for the detection by interferometers in space.

\begin{figure}[t]
  \centering
    \includegraphics[width=0.36\textwidth]{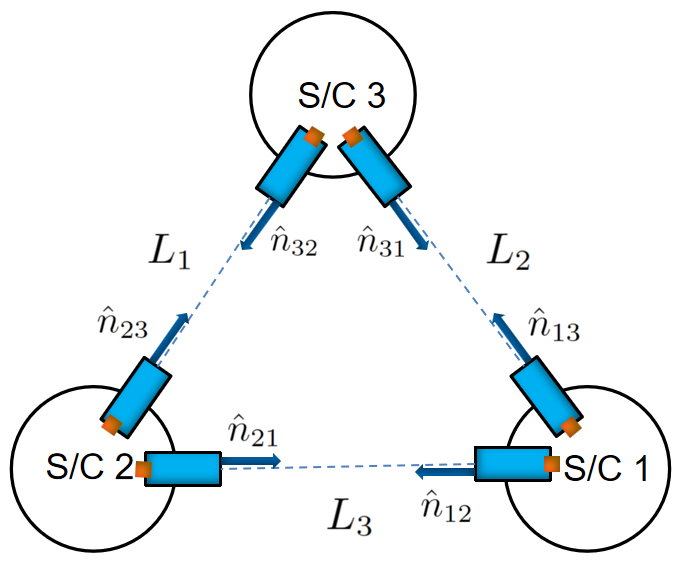}
  \caption{A Schematic of the triangle constellation for LISA-like GW laser interferometers in space. Each spacecraft (S/C) hosts two test masses (orange cubic) and each single-link measurement encodes the distance between two test masses at two ends of an edge. $\hat{n}_{ij}$ denotes the unit direction vector pointing from S/C $i$ to S/C $j$. The arm lengths $L_i$ are changing over time with percent-level variation.}
  \label{spacon}
\end{figure}

Now we derive the signal response of a space-based GW interferometer to the metric perturbations sourced by ULDM. We consider the constellation with three spacecrafts (S/C) as Fig.~\ref{spacon}, which is adopted in most space-based GW detectors, including LISA~\cite{amaroseoane2017laser}, Taiji~\cite{10.1093/nsr/nwx116}, Tianqin~\cite{Luo_2016}, BBO~\cite{PhysRevD.72.083005}, DECIGO~\cite{Kawamura_2006}, $\mu$Ares~\cite{Sesana_2021}, LISAmax~\cite{martens2023lisamax}, and ASTROD-GW~\cite{NI_2013}. The signal is encoded in the fractional frequency shift,
\begin{equation}
 y_{rs}(t)\equiv \frac{\delta \nu_{r s}(t)}{\nu_0}\simeq z_{\Phi}(t)+z_{h}(t) + N(t),
\end{equation}
which parameterizes various effects on the single-link frequency shift $\delta \nu_{r s}$ of the transmitted laser light from S/C $s$ to S/C $r$, including instrumental noises $N(t)$ and geodesic deviations from perturbations $\Phi$ and $h_{ij}$, $z_{\Phi}(t)+z_{h}(t)$. The central laser frequency is taken as $\nu_0 \simeq 2.81\times 10^{14}$~Hz as the wavelength of laser light is $1064$~nm.

\noindent {\bf Instrumental noises and TDI combinations.}---%
Three main instrumental noises, contributing to $\delta \nu_{r s}(t)$, would limit the detection sensitivity, including frequency noise of lasers $p_i(t)$, acceleration noise of test masses $s_\text{acc}$, and noise of optical metrology system $s_\text{oms}$. Frequency noise is related to how monochromatic the laser light is, while acceleration noise denotes the deviation of free fall for test mass, and metrology noise represents the displacement uncertainty of optical interferometers. 

\begin{figure}[t]
  \centering
    \includegraphics[width=0.35\textwidth]{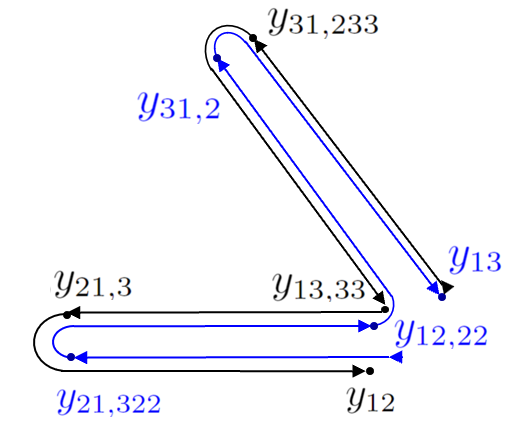}
  \caption{The schematic of Michelson combination $X$ centered at S/C 1. For time delays, we use the convention, $y_{ij,kl\cdots}=y_{ij}(t-L_k-L_l\cdots)$. The light paths of two colors are nearly equal so that laser frequency noise is effectively suppressed.}
  \label{fig:X}
\end{figure}

Due to the mismatch of time-changing arm lengths, the laser frequency noise in a single-link data stream $y_{rs}(t)$, $p_s(t-L_j)-p_r(t)$, does not cancel and dominates over all other ones by several orders. To suppress laser noise to a negligible level, the usual strategy is to postprocess the acquired data by time-delay interferometry (TDI) algorithm~\cite{Tinto:2020fcc}, which makes time-shifts and linear combinations of $y_{rs}(t)$, or equivalently constructs virtual equal-arm interferometry. One example for Michelson combination $X(t)$ can be written as follows,
\begin{align}\label{eq:tdix}
   X(t)=\big{[}&y_{13}(t)+y_{31}(t-L_2)+y_{12}(t-2L_2)\nonumber\\
   &+y_{21}(t-L_3-2L_2)\big{]}-\big{[}y_{12}(t)+y_{21}(t-L_3)\nonumber\\
   &+y_{13}(t-2L_3)+y_{31}(t-L_2-2L_3)\big{]}.
\end{align}
The combinations in two brackets can be geometrically described in Fig.~\ref{fig:X} by the blue and black light paths, respectively. Two paths have almost equal lengths, $2L_2+2L_3$, but in different directions, indicated by the arrows. Many other combinations are possible (see Supplemental Material). The laser frequency noise in such combinations can be suppressed and hence neglected in further discussions. Also $L_i$ appearing in all other contributions can be approximately taken as $L_i=L$, the nominal arm length of the triangle constellation. 

\begin{table*}[t]
\centering
\begin{tabular}{ccccccccc}
  \toprule
   &LISA & Taiji  & Tianqin & BBO & DECIGO & $\mu$Ares & LISAmax & ASTROD-GW  \\
  \midrule
 $L$\,($10^{9}$m) &$2.5 $    & $3$  & $0.17$ &$0.05$&$1\times10^{-3}$ &$395$ &$260$ &$260$ \\
 $s_\text{acc}$\,($10^{-15}\frac{\textrm{m/s}^{2}}{\sqrt{\textrm{Hz}}}$) &$3$  & $3$   & $1$&$3\times 10^{-2}$ & $4\times 10^{-4}$ & $1$ & $3$ & $3$\\
 $s_\text{oms}$\,($10^{-12}\frac{\textrm{m}}{\sqrt{\textrm{Hz}}}$)& $15  $ & $8 $ & $1 $ &$1.4\times 10^{-5}$& $2\times 10^{-6}$ & $50$ &$15$ &$100$ \\
 $\dfrac{\rho}{\rho_0}$($5.0\times 10^{-19}\text{eV}$) & $7.95\times10^2$&$6.53\times 10^2$&$3.82\times 10^3$&$2.00\times10^2$&$1.34\times10^2$&$0.44$&$7.67$&$4.32$\\
  \bottomrule
\end{tabular}
\caption{Arm lengths and instrumental noises of several planned laser interferometers in space. In the last row we give the sensitivities on vector ULDM with mass $5.0\times 10^{-19}\text{eV}$. Here $L$ is the nominal arm length of triangle constellation, while $s_\text{acc}$ and $s_\text{oms}$ are the acceleration noise of test mass and noise from optical metrology system, respectively. Note that LISA/LISAmax/Taiji/Tianqin adopt frequency-dependent noise power spectra~\cite{Babak:2021mhe}, $s^2_\text{acc} \propto [ 1+\left( 0.4 \times 10^{-3}~\textrm{Hz}/f  \right)^{2} ] [ 1+ \left({f}/{8 \times 10^{-3}~\textrm{Hz}} \right)^{4} ] \;$ and $s^2_\text{oms}\propto 1 + \left({2 \times 10^{-3}~\textrm{Hz}}/{f} \right)^{4} $.  \label{tab:pra}}
\end{table*}

On the other hand, acceleration noise and metrology noise cannot be suppressed and consequently limit the detection sensitivity. In Table.~\ref{tab:pra}, we list the noise power spectral density (PSD) for several future laser interferometors in space. To compare with the signal strain in $X$ directly, we calculate the noise power spectrum in Michelson combination, Eq.~\ref{eq:tdix},
\begin{eqnarray}\label{eq:noisepsd}
    N_X(\omega) = 16\sin^{2}\omega L \left[ 2(1+\cos^2\omega L )\frac{s^2_\text{acc}}{\omega^2}  + \omega ^{2}s^2_\text{oms} \right],
\end{eqnarray}
where $\omega$ is the angular frequency and $\omega=2\pi f$. It is expected that acceleration noise would dominate at low frequencies due to the $\omega^{-2}$ factor while optical metrology noise dominates at high frequencies because of $\omega^{2}$ term. In the low-frequency limit, $\omega L \ll 1$, we have $N_X(\omega) \simeq 64L^2s^2_\text{acc}$, which is frequency-independent for constant $s_\text{acc}$. At first sight, it seems longer arms would make detectors less sensitive. However, as we shall shown, the signal PSD also depends on $L$ and eventually prefers larger arm lengths.

\noindent {\bf Sensitivities to ULDM.}---%
To estimate the sensitivity of a laser interferometer to ULDM, we calculate the signal PSD of $X(t)$ in frequency domain, and compare it with the noise PSD, $N_X(\omega)$ in Eq.~\ref{eq:noisepsd}. The ULDM PSD of $X(t)$ in observation duration $T$ is related to the amplitude in frequency domain, $\Tilde{X}(\omega)$, 
\begin{equation}\label{eq:Xpsd}
    S_{X}(\omega)=|\Tilde{X}(\omega)|^{2}/T=S^{j}_\Phi + S^{j}_h,
\end{equation}
where we have separated the contributions from scalar and tensor metric perturbations into $S^{j}_\Phi$ and $S^{j}_h$, respectively, and performed the sky average over the propagation direction of ULDM, and additional average over the polarizations for vector ULDM. We have neglected the cross term $S_{\Phi h}$, which either vanishes after averaging over directions and polarizations, or is subdominant. Parameterizing ULDM as superpositions of plane waves with frequency dispersion $\delta \omega \sim mv^2$, we observe that $S_{X}(\omega)$ has a nearly monochromatic component at $\omega = 2m$~\footnote{Here we focus on the deterministic monochromatic signal, other than the stochastic continuous component at lower frequencies~\cite{Kim:2023pkx}, which has no boost factor $T$ in PSD if there is no an additional detector for correlation analysis.}.

Specifically, for ULDM propagating along some direction $\hat{k}=\Vec{k}/|\Vec{k}| (\Vec{k}=m\Vec{v})$, we have 
\begin{align}
    S^{j}_{\Phi}( \hat{k}) & \simeq 
    v^2_{\text{osc}}T\Big{|}\hat{k}\cdot \hat{n}_{13}\big{[}1-e^{-i8mL}-2e^{-i(2mL+2\Vec{k}\cdot \hat{n}_{13}L)} \nonumber\\
    &+2e^{-i(6mL+2\Vec{k}\cdot \hat{n}_{13}L)}\big{]} -\hat{k}\cdot \hat{n}_{12}\big{[}1-e^{-i8mL} \nonumber\\
    &-2e^{-i(2mL+2\Vec{k}\cdot \hat{n}_{12}L)}+2e^{-i(6mL+2\Vec{k}\cdot \hat{n}_{12}L)}\big{]}\Big{|}^2,
\end{align}
where $v_{\text{osc}}=\kappa^{2}\rho v/4m^{2}$, $\kappa \simeq  \sqrt{4\pi G}$. And
\begin{align}
    S^{s}_{h}( \epsilon_{ij}) \simeq  \frac{64}{9}\kappa^4\rho^2v^4L^4T[(\hat{n}_{12}^{i}\hat{n}_{12}^{j}-\hat{n}_{13}^{i}\hat{n}_{13}^{j})\epsilon_{ij}]^2,
\end{align}
where $\epsilon_{ij}$ is the polarization tensor and should be averaged in later numerical evaluations. 
A quick estimation shows the average of the term in bracket gives an $\mathcal{O}(1)$ factor. 

Under the low-frequency approximation, $m L \ll 1$, we average the propagation direction $\hat{k}$ analytically,
\begin{align}\label{eq:s0}
    S^{s}_{\Phi} \simeq  \frac{64}{15} \kappa^4\rho^2v^2L^4T \; \left[ v^2\sin^2\gamma + 5  m^2L^2\sin^2\frac{\gamma}{2}\right],
\end{align}
where $\gamma$ is the angle between two arms of the triangle, $\gamma \simeq \pi/3$. When $ m L \gg v^2 \simeq 10^{-6}$, the second term in bracket would be dominant and $S^{s}_\Phi \propto m^2v^2L^6$. Otherwise, $S^{s}_\Phi \propto v^4L^4$ and comparable to $S^{s}_h$. In both cases, longer arms give larger signal PSD. 

However, for vector ULDM the dominant contribution comes from the tensor perturbation,
\begin{align}
    S^{v}_{h}( \epsilon_{ij}) \simeq  \frac{256}{9}\kappa^4\rho^2L^4T[(\hat{n}_{12}^{i}\hat{n}_{12}^{j}-\hat{n}_{13}^{i}\hat{n}_{13}^{j})\epsilon_{ij}]^2.
\end{align}
Note that $S^{v}_h$ is significantly larger than $S^{v}_\Phi$, at least by a factor of $1/v^2\sim 10^6$. Therefore $S^{v}_\Phi$ can be neglected. 


It is now evident that larger arm length $L$ gives larger power spectrum $S_X$, which suggests that at the same noise level GW detectors in space with longer arms would have better sensitivity to ULDM. To estimate the sensitivity quantitatively, we define it as the value of density $\rho$ with which the signal-to-noise ratio (SNR) reaches unity,
\begin{equation}\label{eq:snr}
    \text{SNR} = \frac{S_X(\omega)}{N_X(\omega)} = 1.
\end{equation}
At low frequencies we have $\rho \propto \kappa^2 s_\text{acc}/L\sqrt{T}$. When the sensitivity $\rho$ of some detector is less than or equal to DM's local density $\rho_0\simeq 0.4~\text{GeV/cm}^{3}$, namely $\rho/\rho_0\lesssim 1$, such a detector would be able to probe local ULDM in solar system.

\begin{figure}[t]
\centering
    \includegraphics[width=0.5\textwidth]{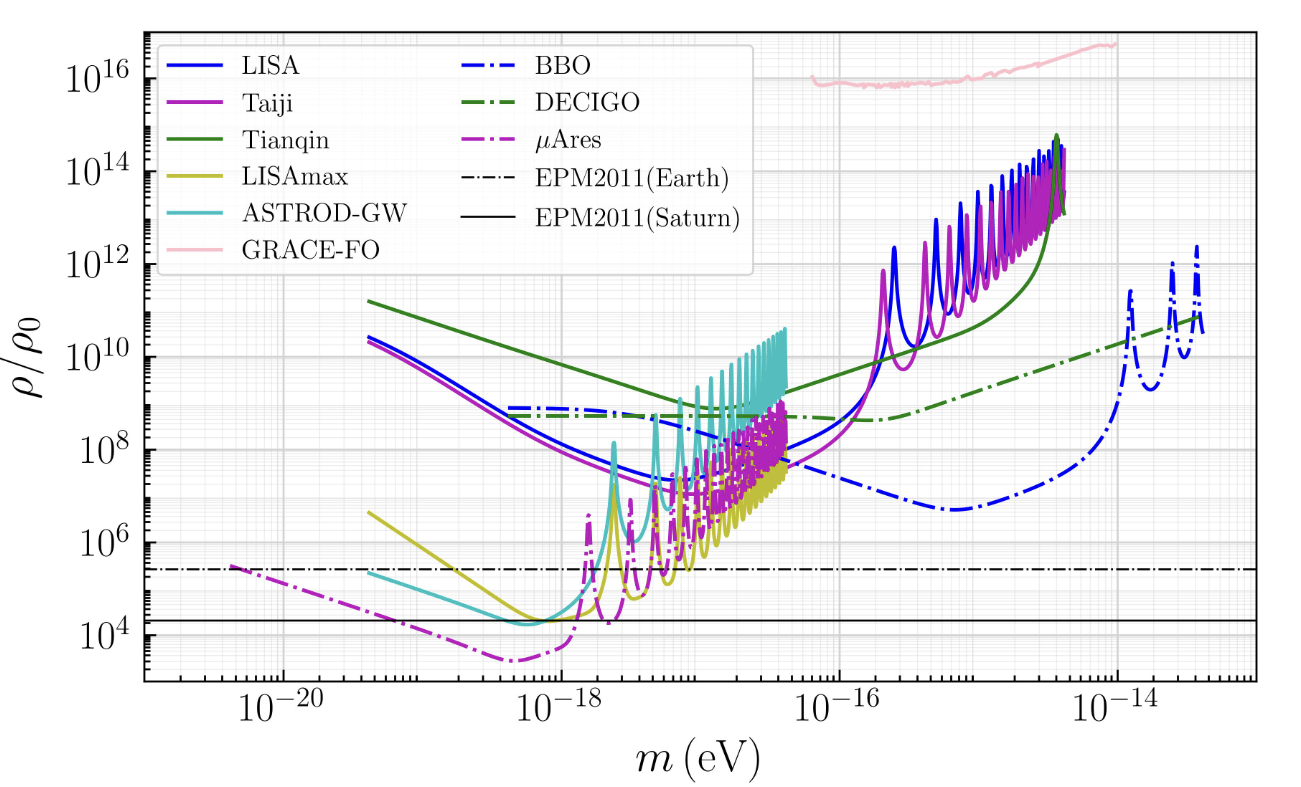}
    \includegraphics[width=0.5\textwidth]{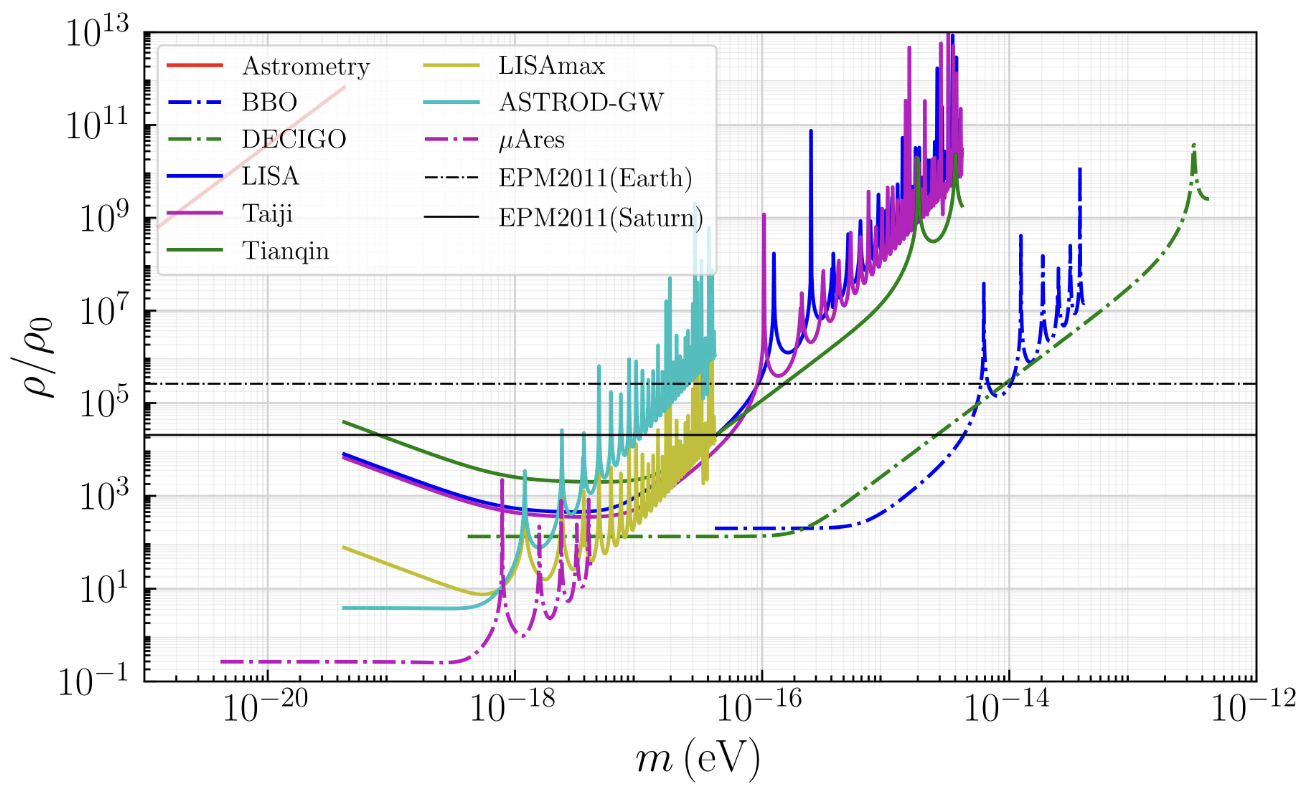}
  \caption{(Upper) Sensitivities to $\rho$ for scalar ULDM. (Lower) Sensitivities to $\rho$ for vector ULDM. We assume one-year observation time for LISA, Taiji, Tianqin, BBO, DECIGO, LISAmax and ASTROD-GW, ten years for $\mu$Ares and 13 days for GRACE-FO (pink curve). For comparison, we also show the constraints (horizontal lines) derived from the solar system planetary ephemerides \cite{Pitjev_2013} and the estimated astrometric detection threshold (the red line in lower panel).}
  \label{fig:sensi}
\end{figure}

In Fig.~\ref{fig:sensi} we plot the sensitivities of various GW detectors to scalar and vector ULDM. For scalar ULDM, the most sensitive range is around $m\sim 10^{-18}$~eV by $\mu$Ares~\cite{Sesana_2021}, which is aiming at detecting $\mu$Hz GWs. The sensitivity surpasses that from planetary ephemerides in solar system~\cite{Pitjev_2013}, reaching $\rho/\rho_0 \sim 10^3$. This large value also suggests that, unless there are huge improvements on instrumental noises by several orders of magnitude or there are DM clumps near solar system, it is difficult to probe local scalar ULDM gravitationally through GW detectors in space. 

Interestingly, for vector ULDM with $m \lesssim 10^{-18}$~eV $\mu$Ares is able to probe $\rho/\rho_0 \simeq 0.4$, indicating the possibility to detect local DM gravitationally. In this case, at low frequencies the flat behaviour of sensitivities from $\mu$Ares, ASTROD-GW, DECIGO and BBO as well, results from the frequency independence of $N_X(\omega)$ and $S_h$ when $\omega L\ll 1$. For LISA, Taiji, Tianqin and LISAmax, the adopted acceleration noises $s_\text{acc}$ have non-trivial frequency dependence, making them less sensitive at lower frequencies. The plot demonstrates that at low frequencies the sensitivity goes as $\rho \propto L^{-1}$, which is quite different from the scalar ULDM case where $\rho \propto L^{-2}$ for $1\gg mL\gg v^2\sim 10^{-6}$, indicated by Eq.~\ref{eq:s0}. It also exhibits that all discussed GW detectors in space would surpass planetary ephemerides in some mass ranges. These findings extends the scientific objects for future GW detectors in space. 

We have conducted the above numerical analysis by employing the first-generation time-delay interferometry (TDI), which assumes static arm lengths. However, the sensitivities and conclusions do not change if we use the second-generation TDI with time-changing arm lengths, which doubles the terms in $X(t)$ and the light paths in Fig.~\ref{fig:X}. The resulting noise and signal PSDs would be modified by the same factor $4\sin^2(2mL)$ and cancel out in the signal-to-noise ratio. Furthermore, we have only used the Michelson channel. Many other channels are possible, which can provide additional information of the signal, like the spin of ULDM (Supplemental material). 

For comparison, we show the inferred limit from the GRACE-FO satellite project~\cite{PhysRevLett.123.031101}, which has validated the laser metrology to be deployed in LISA. There are only two spacecrafts in GRACE-FO whose separation is monitored by laser interferometers. Although not capable to detect GW or ULDM, the overall noise level in the collected data can still give a constraint, shown as the pink curve in the upper plot. Similar to GWs, the tensor metric perturbation induced by ULDM also leads to a distortion of the apparent angular positions of stars, a rough estimation for its detectability with Gaia-like astrometric accuracy (see Supplemental Material) is shown in the lower plot. Both limits are considerably weaker than that from planetary ephemerides, and GW detectors in space. 

\noindent {\bf Conclusion and Outlook.}---%
We have elucidated that future space-based GW laser interferometers can provide new insights into detecting ULDM and extend the scientific objectives. Without assuming any additional interactions between ULDM and normal matter, the gravitational perturbations caused by vector ULDM with mass $m \lesssim 10^{-18}~$eV in solar system is able to perturb the geodesics of test masses at a detectable level if the arm length of the triangle constellation is comparable to the diameter of Mars' orbit. For scalar ULDM, although not capable to make a detection, GW detectors would probe the mass range around $10^{-18}~$eV that other experiments are challenge to explore. This investigation highlights the unique advantage of GW detectors in space. Detection of such a signal would directly suggest the wave nature of DM and its mass and spin, indicating the fundamental theories of DM. 

To distinguish the ULDM signal from GW events from compact binaries nearby galaxies, it is necessary to go beyond the power spectral density and perform matched filtering in time domain or even joint observation with multiple detectors. A more sophisticated treatment would be comparing the outcome of various TDI combinations, which at low frequencies have different sensitivities to ULDM of different spins and GWs. The anisotropic distribution of such GWs would also be distinct. Dedicated investigations are warranted in future work.

\section*{acknowledgement}
Y.T. is supported by the National Key Research and Development Program of China under Grant No.2021YFC2201901, the National Natural Science Foundation of China (NSFC) under Grant No.~12347103 and the Fundamental Research Funds for the Central Universities. Y.L.W. is supported by the National Key Research and Development Program of China under Grant No.2020YFC2201501, and NSFC under Grants No.~11690022, No.~11747601, No.~12347103, and the Strategic Priority Research Program of the Chinese Academy of Sciences under Grant No. XDB23030100.
\bibliography{main}

\clearpage
\onecolumngrid

\begin{center}
  \textbf{\large Supplementary Material for Detecting ULDM with Laser Interferometers in Space}\\[.2cm]
    \vspace{0.05in}
  {Jiang-Chuan Yu, Yan Cao, Yong Tang, and Yue-Liang Wu}
\end{center}

\setcounter{equation}{0}
\setcounter{figure}{0}
\setcounter{table}{0}
\setcounter{section}{0}
\setcounter{page}{1}
\makeatletter
\renewcommand{\theequation}{S\arabic{equation}}
\renewcommand{\thefigure}{S\arabic{figure}}
\renewcommand{\theHfigure}{S\arabic{figure}}%
\renewcommand{\thetable}{S\arabic{table}}

\onecolumngrid

This Supplementary Material contains supporting analysis, formulas and figures for the main letter. In Sec.~\ref{metric}, we show the specific form of metric perturbations induced by scalar and vector ULDM. In Sec.~\ref{response}, we present the single-link laser signal of purely gravitational ultralight DM fields for space-based interferometers. We find that the signal consists of three kinds of components, namely the frequency shift of the photons induced by scalar and tensor metric perturbations and the motions of test-masses. Then in Sec.~\ref{sensitivity} we derive the signal power spectrum and projective sensitivities of various space-based GW interferometers to the local ULDM for TDI-X channel and compare the responses of various TDI combinations. Finally, we show the DM signals given by other experiments in Sec.~\ref{other}, especially other laser interferometers in space and astrometric effects.  

\section{Metric perturbations from ultralight dark matter}\label{metric}
In this section, we review the metric perturbations sourced by nonrelativistic ULDM of various spins in detail, focusing on the case of real fields. In the Poisson gauge defined by $\partial_i h_{ij}=\delta_{ij}h_{ij}=0$, the metric perturbations is given by the linearized Einstein equations \cite{Nomura:2019cvc},
\begin{align}
	\partial_i \partial^i \Phi&=4 \pi G\,T_{00},\label{Einstein1}
\\
	3 \ddot{\Phi}+\partial_i \partial^i(\Psi-\Phi)&=4 \pi G\, T_k^k,\label{Einstein2}
\\
	\ddot{h}_{i j}&=16 \pi G\,\left(T_{i j}-\frac{1}{3} \delta_{i j} T_k^k\right),\label{Einstein3}
\end{align}
with $T_{\mu \nu}$ being the energy-momentum tensor of ULDM, in Eq.~\eqref{Einstein3} we have neglected the spatial gradients of all metric perturbations (namely $\Phi,\Psi,h_{ij}$), since they are highly suppressed in the nonrelativistic limit of ULDM.

\subsection{Scalar field}
A real massive scalar field is described by the Klein-Gordon Lagrangian
\begin{equation}
	\mathcal{L}=-\frac{1}{2}(\partial \phi)^2-\frac{1}{2}m^2\phi^2,
\end{equation}
with its energy-momentum tensor given by
\begin{equation}
	T_{a b}=\partial_a \phi \partial_b \phi-g_{a b}\left(\frac{1}{2} g^{c d} \partial_c \phi \partial_d \phi+\frac{1}{2}m^2\phi^2\right).
\end{equation}
Treating DM as a plane wave, we have
\begin{equation}
	\phi(t,\vec x)=\frac{\sqrt{2\rho}}{m}\cos(\omega t-\vec k \cdot \vec x),
\end{equation}
with $\omega=\sqrt{|\vec k|^2+m^2}$. Using Eqs.~\eqref{Einstein1}, \eqref{Einstein2} and \eqref{Einstein3}, we obtain the nonzero oscillating part of the metric perturbations as
\begin{align}
    \Psi_c(t,\vec x)&=\Psi_0^s\cos(2\omega t-2\vec k \cdot \vec x), \Psi_0^s =-\pi G\frac{\rho}{m^2}=-6.6\times10^{-26}\left(\frac{\rho}{0.4\text{GeV/cm}^{3}}\right)\left(\frac{10^{-18}\text{eV}}{m}\right)^{2},
    \\
    \Phi_c(t, \vec x)&=\Phi_0^s\cos(2\omega t-2\vec k \cdot \vec x), \Phi_0^s=\pi G\frac{\rho}{m^2}=6.6\times10^{-26}\left(\frac{\rho}{0.4\text{GeV/cm}^{3}}\right)\left(\frac{10^{-18}\text{eV}}{m}\right)^{2},
    \\
    h_{ij}(t, \vec x)&=h_0^s\cos(2\omega t-2\vec k \cdot \vec x)\epsilon_{ij}^s,
    h_0^s=\frac{4\pi G}{3}v^2\frac{\rho}{m^2}=2.55\times10^{-31}\left(\frac{\rho}{0.4\text{GeV/cm}^{3}}\right)\left(\frac{10^{-18}\text{eV}}{m}\right)^{2}, 
\end{align}
where $\hat{k}$ is the unit propagation vector and $ \epsilon_{ij}^s=3\hat{k}_i\hat{k}_j-\delta_{ij}$.

\subsection{Vector field}
A real massive abelian vector field $A_\mu$ is described by the Proca Lagrangian:
\begin{equation}
	\mathcal{L}=-\frac{1}{4}F_{ab}F^{ab}-\frac{1}{2}m^2A_aA^a.	
\end{equation}
with its energy-momentum tensor given by
\begin{equation}
	T_{a b}=-\frac{1}{4} g_{a b}F_{cd}F^{cd}+g^{c d} F_{a c} F_{b d}+m^2\left(A_a A_b-\frac{1}{2} g_{a b} A_cA^c\right).
\end{equation}
Similarly, we treat the vector field as
\begin{equation}
	A_i(t,\vec x)=\alpha_i\frac{\sqrt{2\rho}}{m}\cos(\omega t-\vec k \cdot \vec x),
\end{equation}
where the polarization vector $\vec \alpha=(\sin\theta \cos\phi,\sin\theta \sin \phi, \cos \theta)$ is a unit vector along the angular direction $(\theta,\phi)$ of oscillation. Using Eqs.~\eqref{Einstein1}, \eqref{Einstein2} and \eqref{Einstein3} we obtain the nonzero oscillating part of the metric perturbations as
\begin{align}\label{vector_metric}
\Phi_c(t, \vec x)&=\Phi_0^v\cos(2\omega t-2\vec k \cdot \vec x), \Phi_0^v=\frac{1}{3}\pi G\frac{\rho}{m^2}=2.2\times10^{-26}\left(\frac{\rho}{0.4\text{GeV/cm}^{3}}\right)\left(\frac{10^{-18}\text{eV}}{m}\right)^{2},
\\
h_{ij}(t, \vec x)&=h_0^v\cos(2\omega t-2\vec k \cdot \vec x)\epsilon_{ij}^v,
h_0^v=\frac{8}{3}\pi G\frac{\rho}{m^2}=1.7\times10^{-25}\left(\frac{\rho}{0.4\text{GeV/cm}^{3}}\right)\left(\frac{10^{-18}\text{eV}}{m}\right)^{2}, 
\end{align}
where $\epsilon_{ij}^v=\delta_{ij}-3\alpha_i\alpha_j$. Here we omitted $\Psi$ because its magnitude is comparable to $\Phi$ and the signal generated by $\Psi$ is negligible compared to the signal generated by $h_{ij}$.

\section{Detector response}\label{response}
The one-way or single-link laser signal, sent from S/C $s$ at $\Vec{x}_s$ and received by S/C $r$ at $\Vec{x}_s$, is affected by the metric perturbations and usually formulated as the fractional frequency shift over the central value $\nu_0\simeq 2.81\times 10^{14}$~Hz,
\begin{equation}\label{sinfre}
 y_{rs}(t)=\frac{\delta \nu_{r s}(t)}{\nu_0}\equiv\frac{\nu_{r s}(t)-\nu_0}{\nu_0}
 =z_{\Phi}(t)+z_{h}(t)-\frac{d \delta t_{r s}(t)}{d t},
\end{equation}
where each contribution can be calculated as (see for example \cite{Nomura:2019cvc}),
\begin{align}\label{z}
z_{\Phi}(t) &= \Phi(t,\Vec{x}_{r}) - \Phi(t-L,\Vec{x}_{s}) +\int_{t-L}^{t} d t'\, n^i_{rs} \partial_i(\Psi+\Phi),\nonumber\\
z_{h}(t) &= \frac{1}{2}n_{rs}^{i}n_{rs}^{j}[h_{ij}(t-L,\Vec{x}_{s})-h_{ij}(t,\Vec{x}_{r})]+\int_{t-L}^tdt'\,n^k_{rs}\partial_k \left(-\frac{1}{2} n^i_{rs} n^j_{rs} h_{i j}\right), \nonumber\\
\delta t_{r s} &=-\hat{n}_{rs} \cdot\left[\delta \Vec{x}\left(t, \Vec{x}_r\right)-\delta \Vec{x}\left(t-L, \Vec{x}_s\right)\right]. 
\end{align}
Here $\delta \vec{x}$ is the displacement of test mass induced by the metric perturbation, given by the acceleration \cite{Costa:2012cw}
\begin{equation}
\frac{d^2\vec x}{dt^2}=-\nabla\Psi -\upsilon_m^j\partial_t\left[h_{ij}-(2\Phi+\Psi)\delta_{ij}\right],
\end{equation}
which in the nonrelativistic limit is dominated by the first term, since the contributions from $\Phi$ and $h_{ij}$ are suppressed by the test mass velocity $\vec v_m = d \vec x/dt$. In the case of scalar field, the gravitational acceleration is given as usual by $-\nabla \Psi$. For vector or tensor fields, the effect from test mass motion (being nonrelativistically suppressed) is negligible comparing with the gravitational modulation of photon propagation given by Eq.~\eqref{sinfre}.

As displayed in Eq.~\eqref{z}, both $z_{\Phi}$ and $z_{h}$ can be separated into two parts: a dominant part involving only the time derivatives of the bosonic field and a sub-dominant part which is nonrelativistically suppressed by the velocity dispersion of the dark matter $v$. Meanwhile, the frequency shift term induced by detector motion ($d \delta t_{r s}/d t$) is also suppressed by $v$. Since as ULDM the bosonic field under consideration is highly nonrelativistic, for the calculations we adopt the convenient approximation $\omega\approx 2m$.

\section{Signal power spectrum and sensitivities of TDI combinations}\label{sensitivity}
\subsection{Signal power spectrum}
For the static and equal arm length case, other than $X$ channel in Eq.~\ref{eq:tdix}, the Michelson $Y$ and $Z$ combinations can be obtained by the cyclic permutation of the indices ($1\rightarrow2\rightarrow3\rightarrow1$). 
$X$, $Y$ and $Z$ can be further assembled into three optimal combinations, $A, E$ and $T$~\cite{Prince:2002hp},
\begin{eqnarray}\label{tdia}
    A(t) &=& \frac{1}{\sqrt{2}} \left[ Z(t)-X(t) \right],\nonumber\\
    E(t) &=& \frac{1}{\sqrt{6}} \left[ X(t)-2Y(t)+Z(t) \right],\nonumber\\
    T(t) &=& \frac{1}{\sqrt{3}} \left[ X(t)+Y(t)+Z(t) \right].
\end{eqnarray}

We shall illustrate with the $X$ channel and calculate the signal power spectrum. Substitute Eq.~\eqref{sinfre} into Eq.~\eqref{eq:tdix}, we obtain
 \begin{equation}\label{tdi}
\begin{aligned}
   X(t)&=\frac{1}{2}(n_{12}^{i}n_{12}^{j}-n_{13}^{i}n_{13}^{j})[h_{ij}(t,\Vec{x}_{1})-2h_{ij}(t-2L,\Vec{x}_{1})+h_{ij}(t-4L,\Vec{x}_{1})]\\
   &\quad +\hat{n}_{13} \cdot\frac{\mathrm{d}}{\mathrm{d} t}[\delta \Vec{x}(t, \Vec{x}_1)-2\delta \Vec{x}(t-L, \Vec{x}_3)+2\delta \Vec{x}(t-3L, \Vec{x}_3)-\delta \Vec{x}(t-4L, \Vec{x}_1)]\\
   &\quad -\hat{n}_{12} \cdot\frac{\mathrm{d}}{\mathrm{d} t}[\delta \Vec{x}(t, \Vec{x}_1)-2\delta \Vec{x}(t-L, \Vec{x}_2)+2\delta \Vec{x}(t-3L, \Vec{x}_2)-\delta \Vec{x}(t-4L, \Vec{x}_1)],
\end{aligned}
\end{equation}

Note that the total contribution form $z_\Phi$ cancels in $X$ channel.  Then in frequency domain the signal has amplitude
\begin{equation}
\Tilde{X}(\omega) = \int_{0}^{T} X(t) e^{-i\omega t}dt=\Tilde{X}_{\Phi}(\omega)+\Tilde{X}_{h}(\omega),
\end{equation}
with
\begin{equation}
\begin{aligned}
    \tilde X_\Phi(m)&=v_{\text{osc}}\delta(\omega-2m)\lbrace(\hat{k}\cdot \hat{n}_{13})[1-2e^{-i(2mL+2\Vec{k}\cdot \hat{n}_{13}L)}+2e^{-i(6mL+2\Vec{k}\cdot \hat{n}_{13}L)}-e^{-i8mL}]\\
    &\quad-(\hat{k}\cdot \hat{n}_{12})[1-2e^{-i(2mL+2\Vec{k}\cdot \hat{n}_{12}L)}+2e^{-i(6mL+2\Vec{k}\cdot \hat{n}_{12}L)}-e^{-i8mL}]\rbrace,\\
    &=v_{\text{osc}}T\lbrace(\hat{k}\cdot \hat{n}_{13})[1-2e^{-i(2mL+2\Vec{k}\cdot \hat{n}_{13}L)}+2e^{-i(6mL+2\Vec{k}\cdot \hat{n}_{13}L)}-e^{-i8mL}]\\
    &\quad-(\hat{k}\cdot \hat{n}_{12})[1-2e^{-i(2mL+2\Vec{k}\cdot \hat{n}_{12}L)}+2e^{-i(6mL+2\Vec{k}\cdot \hat{n}_{12}L)}-e^{-i8mL}]\rbrace,\\
    \tilde 
    X_h(m)&=\frac{\delta(\omega-2m)}{2}(n_{12}^{i}n_{12}^{j}-n_{13}^{i}n_{13}^{j})h_{0}\epsilon_{ij}(1-2e^{-i4mL}+e^{-i8mL})\\
    &=\frac{T}{2}(n_{12}^{i}n_{12}^{j}-n_{13}^{i}n_{13}^{j})h_{0}\epsilon_{ij}(1-2e^{-i4mL}+e^{-i8mL}),
\end{aligned}
\end{equation}
where $v_{\text{osc}}=\kappa^{2}\rho|\Vec{k}|/4m^{3}=\kappa^{2}\rho v/4m^{2}$ is the amplitude of the velocity oscillation. In the above calculation, we have taken $\omega = 2m$. Then the ULDM PSD in observation duration is given by 
\begin{align}\label{ss}
    S_X(\omega)=|\tilde{X}(\omega)|^2/T\simeq S_\Phi+S_h.
\end{align}
For scalar ULDM, $S_{\Phi}$ and $S_h$ are at the same magnitude, so we keep both of them. For vector ULDM, $S_h \gg S_{\Phi}$, so we neglect $S_{\Phi}$ for convenience. As for the cross term $S_{\Phi h}$, it will vanish after averaging the polar angles in the case of scalar and vector ULDM. Therefore, we do not include $S_{\Phi h}$ in Eq.(\ref{ss}).

\subsection{Sensitivities on DM density for $X$ channel}
For Michelson $X$ channel, combined with the signal spectrum above and the noise spectrum of instruments, we are able to obtain the expression of the sensitivities on $\rho$. For scalar ULDM, we have
\begin{align}
    \rho(f) &\simeq \frac{1}{8\kappa^{2}vL^2}\frac{1}{\sqrt{\frac{1}{3}m^2L^2\sin^2\frac{\gamma}{2}+(\frac{1}{15}\sin^2\gamma+\frac{1}{9}\overline{|G|^{2}})v^2}}\sqrt{\frac{N_X(f)}{T}}
\end{align}
with the geometry factor $G=(n_{12}^{i}n_{12}^{j}-n_{13}^{i}n_{13}^{j})\epsilon_{ij}$. For different types of fields, the form of $\epsilon_{ij}$ will also vary, and the above line refers to the average of propagation angles or polarization angles. For $mL \gg v$, $(\frac{1}{15}\sin^2\gamma+\overline{|G|^{2}})v^2$ can be ignored. For vector ULDM, we have
\begin{equation}
    \rho(f) \simeq \frac{3m^{2}}{2\kappa^{2}\sqrt{\overline{|G|^{2}}}(1-\cos4mL)}\sqrt{\frac{N_X(f)}{T}}\overset{mL\to 0}{\simeq} \frac{3}{16\kappa^{2}\sqrt{\overline{|G|^{2}}}L^2}\sqrt{\frac{N_X(f)}{T}}.
\end{equation}
For vector ULDM, the first approximately equal sign in the equation represents our use of $S_X(\omega)\simeq S_h$, since $S_\Phi$ and the cross term in the power spectrum are negligible compared to $S_h$. 

\subsection{Comparison of different TDI combinations}
\begin{figure}[!h]
  \centering
    \includegraphics[width=0.48\textwidth]{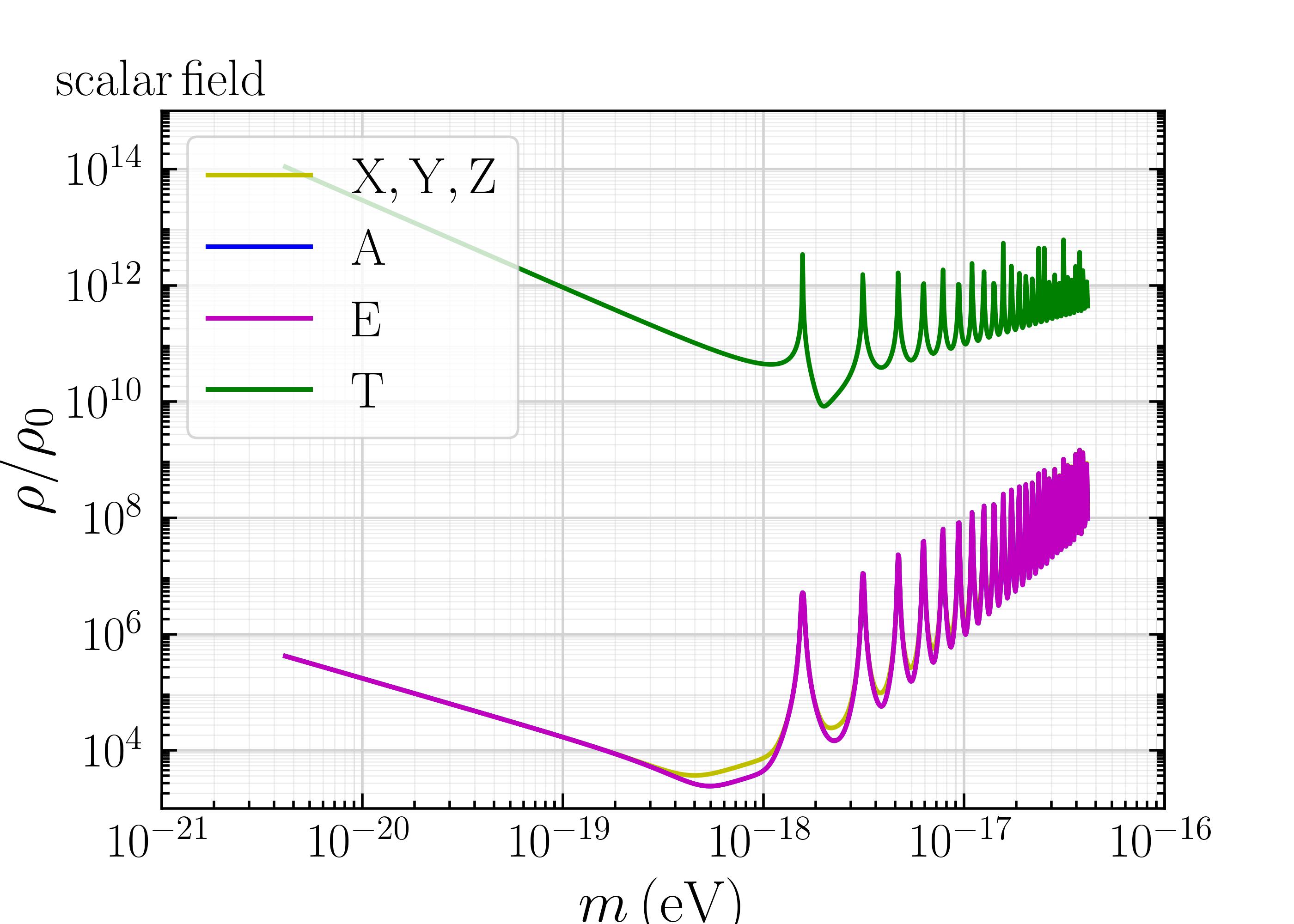}
    \includegraphics[width=0.48\textwidth]{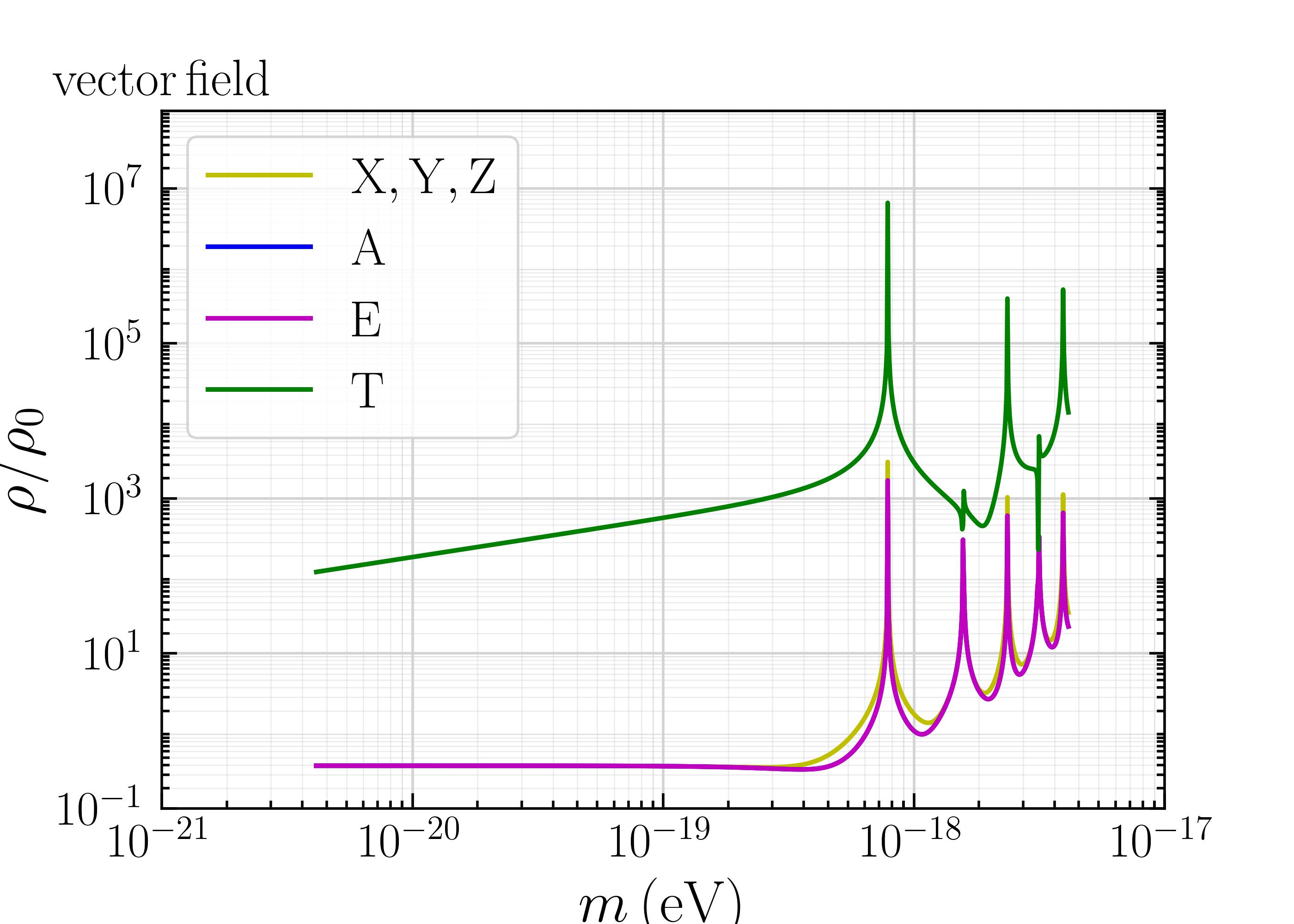}
  \caption{Projected sensitivities on $\rho$ of $\mu$Ares for scalar and vector ULDM. Michelson combinations X, Y, Z and optimal combinations A, E, T are showed. The observation time is 10 yrs.}
  \label{scacon}
\end{figure}

Here we provide a short discussion about the different responses or sensitivities of other combinations. As seen from Fig.~\ref{scacon}, the sensitivities of $A$ and $E$ to ULDM is almost the same as that of $X$. However, $T$ has a very different behavior. Compared to other channels, due to its geometric configuration, the response cancels out in the leading order in $T$~\cite{PhysRevD.108.083007}, resulting in a significant decrease of sensitivities. This suggest that it will provide an effective way to distinguish ULDM of different spins. 

\section{Signals of other experiments}\label{other}
\subsection{Grace Follow-on}
The run data of other laser experiments in space, such as Grace Follow-on (GRACE-FO), can also provide limits on the density of dark matter. The physical principle is very similar. The gravitational perturbation of ULDM will cause the frequency shift of photons and the displacement of the test mass, which will affect the corresponding phases. For GRACE-FO, two satellites separated by $L = 200 $km form a single interferometer with arm-length $2L$, and the half-round displacement noise spectrum of Grace follow-on mission is given by~\cite{PhysRevLett.123.031101}. Then the response is given by 
\begin{equation}
    \begin{aligned}
         y(t)= 2mL^2\Phi_0 \sin2mt-n^in^j\epsilon_{ij}mL^2h_0\sin2mt-vL\Psi_0\hat{k}\cdot \hat{n}\cos2mt,
    \end{aligned}
\end{equation}
and the Fourier amplitude in the finite observation duration $T$ is 
\begin{equation}
    \begin{aligned}
    \tilde
         y(\omega)\simeq (2mL^2\Phi_0 -n^in^j\epsilon_{ij}mL^2h_0-vL\Psi_0\hat{k}\cdot \hat{n})T.
    \end{aligned}
\end{equation}
Then the power spectrum is given by 
\begin{equation}
    \begin{aligned}
        S(\omega) \simeq |2mL^2\Phi_0 -n^in^j\epsilon_{ij}mL^2h_0-vL\Psi_0\hat{k}\cdot \hat{n}|^2 T.
    \end{aligned}
\end{equation}

The first term and the third term are determined by the scalar part of the metric perturbation, and the second term is determined by the tensor part. Correspondingly the constraint on $\rho$ can be derived. For example, for vector ULDM, we have
\begin{equation}
    \rho(f) \simeq \frac{12m^2}{\kappa^2}\sqrt{\frac{N(f)}{|-2mL^2 +8n^in^j\epsilon_{ij}mL^2+vL\hat{k}\cdot \hat{n}|^2}}.
\end{equation}
Considering the difference in coefficients of metric perturbations between scalar and vector fields, the coefficient 12 in the above equation should be replaced with 4 in the case of scalar field. The unit of $N(f)$ here is m/$\sqrt{\text{Hz}}$ , which is different from space-based gravitational wave detectors. The magnitude of the signal is at the same level among scalar, vector and tensor ULDM. 

\subsection{Astrometric effects}
For a stationary observer in the tensor perturbation described by a plane wave $h_{ij}(\xi=-v_pt+\vec n \cdot\vec x)$ with $|\vec n|=1$, the apparent deviation of the angular direction of a light source from its true direction $\vec N$ is given by
\begin{equation}
\Delta N_i=R_{ilj}h_{lj},\quad
R_{ilj}=\frac{\frac12N_lN_j}{1+\vec{n}\cdot\vec{N}/v_p}(N_i+n_i/v_p)-\frac12N_j\delta_{il}.
\end{equation}
here we have extended the result for $v_p=1$ in \cite{Book:2010pf}. For the tensor perturbations sourced by non-relativistic ULDM, we use the approximation $v_p=\infty$, hence
\begin{equation}
	R_{ilj}=\frac12N_lN_jN_i-\frac12N_j\delta_{il}.
\end{equation}
Using the metric perturbation given by Eqs.~\eqref{vector_metric} for vector ULDM, we obtain
\begin{equation}
|\Delta \vec{N}(\vec{N},t)|=A|\cos (2\omega t-2\vec k\cdot \vec x)|,
\quad
A=\frac{4}{3}\pi\frac{\rho}{m^2}|(N_lN_jN_i-N_j\delta_{il})\epsilon_{lj}|.
\end{equation}
An optimistic estimation for the sensitivity of a single astrometric measurement to this oscillating signal is given by $A_\text{max}>10\,\mu\text{as}=4.8\times 10^{-11}\,\text{rad}$ (for Gaia-like accuracy \cite{collaboration2022gaia,Moore:2017ity,Jaraba:2023djs,Mentasti:2023gmr}), where $A_\text{max}$ is the maximum value of $A$ with an optimal choice of field polarization. We can obtain $\rho/\rho_0>5\times 10^{10}\left(\frac{m}{10^{-20}\,\text{eV}}\right)^2$. 

\end{document}